# A Review of Multiscale Thermal Modeling in Heterogeneous 3D ICs

**Baibhari Priya Barua,** *Graduate Student Member*, IEEE, **Md Rahatul Islam Udoy,** *Graduate Student Member, IEEE*, **and Ahmedullah Aziz,** *Senior Member, IEEE*
Department of Electrical Engineering and Computer Science, University of Tennessee, Knoxville, TN 37996 USA.

Corresponding author: Ahmedullah Aziz (e-mail: aziz@utk.edu).

**ABSTRACT** Thermal behavior has become a first-order constraint in advanced 2.5D/3D integrated circuits (ICs) and heterogeneous packages. As power densities rise and multiple active dies are vertically integrated, heat removal paths become constricted, elevating junction temperatures, magnifying temperature gradients, and exacerbating reliability risks. This review synthesizes the physical mechanisms, modeling assumptions, and analysis methods that govern multiscale thermal transport in 3D ICs, with emphasis on interface-dominated conduction, material anisotropy, and strong electrothermal coupling. We unify device-to-system scales into a coherent framework, analyzing trade-offs among compact thermal models (CTMs), finite element/finite difference methods (FEM/FDM), Green's function and semi-analytical techniques, reduced-order and multi-fidelity methods, and physics-informed machine learning (PIML), while highlighting the central role of thermal boundary resistance (TBR) and variability in thermal interface materials (TIMs), the pitfalls of decoupled electrical/thermal analyses, and the need for rigorous validation against measurements. Finally, we outline practical design guidelines and a forward-looking research agenda that integrates physics-based modeling, data-driven surrogates, and in-situ sensing to enable thermally aware co-optimization across the IC–package–system hierarchy.

## I. INTRODUCTION

Heterogeneous 3D integration, spanning 2.5D interposer-based chiplets and true 3D vertical stacking, has emerged as a practical path beyond monolithic system-on-chip scaling for high-performance computing and AI accelerators[1], [2], [3]. By shortening global interconnects and collocating logic with high-bandwidth memory, these architectures increase package-level bandwidth while reducing communication latency and energy per bit. Recent accelerator platforms combining chiplet-based compute dies with stacked HBM demonstrate the practicality of this approach on a scale[3], [4]. In parallel, Intel's Foveros and Foveros Direct technologies illustrate the shift toward fine-pitch hybrid bonding for logic-on-logic and logic-on-memory integration, further narrowing the boundary between on-chip and in-package communication[5]. More broadly, the package has evolved from a passive enclosure into an active integration fabric that sets interconnect density, power delivery, and thermal limits while enabling the co-integration of dies with different functions, technology nodes, and materials.

The same architectural features that make heterogeneous 3D integration attractive also make heat removal substantially more difficult[6], [7]. Vertical stacking (shown in Fig. 1) narrows heat-dissipation pathways, places active tiers farther from direct cooling interfaces, and introduces multiple interfaces whose thermal resistance can strongly influence junction temperature, hotspot formation, and long-term reliability [6], [7], [8]. In addition, temperature is not merely a consequence of power: it can degrade device performance, increase leakage, and intensify Joule heating, thereby creating positive electrothermal feedback that aggravates hotspot severity and, in extreme cases, may trigger thermal runaway [9].

Taken together, the existing review literature remains fragmented. Recent reviews of 3D stacked chips and 3-DHI thermal management primarily emphasize heat-dissipation enhancement, cooling technologies, package/interposer constraints, and other thermal bottlenecks at the chip, package, or system level [6], [10]. By contrast, broader multiscale thermal transport perspectives focus on atomistic-to-continuum heat-transfer physics and computational methods in solids and interfaces, rather than on the cross-scale modeling, handoff, validation, and design-decision workflow required for heterogeneous 3D ICs [11].

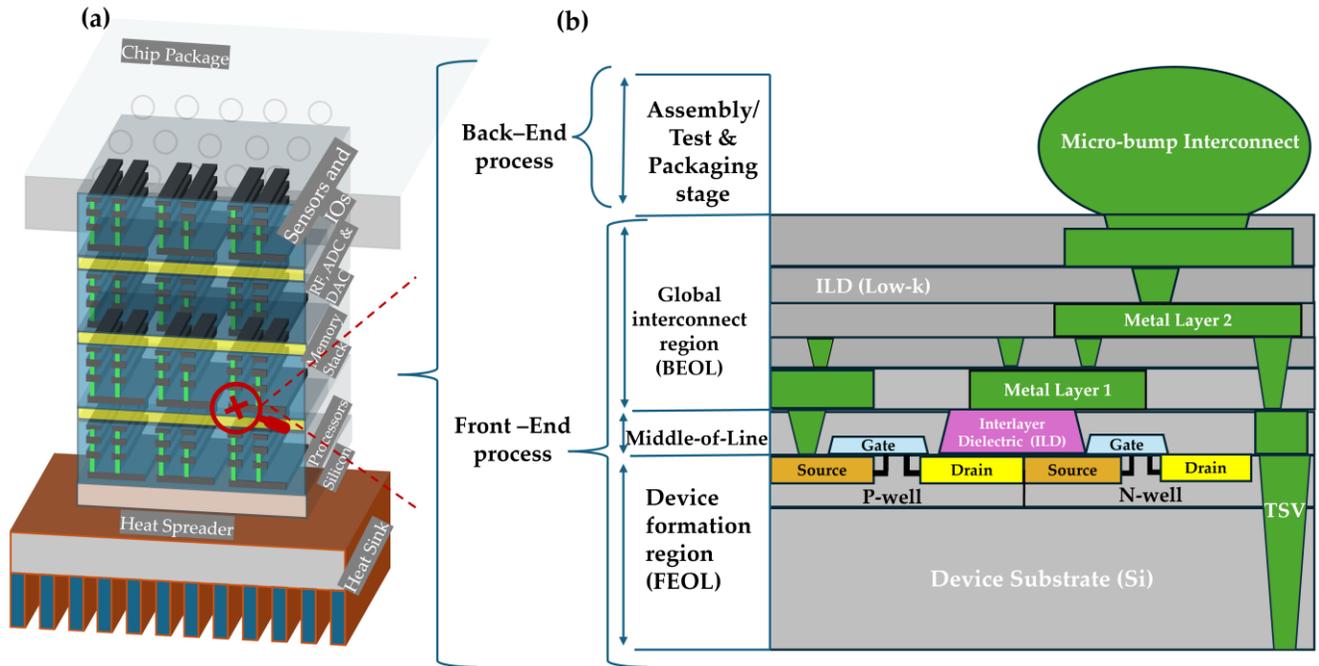

**FIGURE 1.** *Representative heterogeneous 3D stack and multiscale zoom from package, die tiers, and interfaces to BEOL and device-level features.*
(a) System-level schematic of a heterogeneous 3D integrated circuit comprising vertically stacked processor, memory, RF/analog and sensor dies, integrated above a heat spreader and heat sink for thermal management. The architecture highlights functional partitioning across tiers and vertical die-to-die integration within a single package platform.
(b) Cross-sectional magnified view illustrating the hierarchical interconnect stack, including micro-bump or hybrid bonding interfaces, through-silicon vias (TSVs), back-end-of-line (BEOL) metal and interlayer dielectric (ILD) layers, and underlying transistor devices in the front-end-of-line (FEOL). The figure emphasizes the multiscale nature of 3D heterogeneous integration, spanning package-level heat dissipation to nanoscale device and interconnect structures, and underscores the strong electro–thermal coupling across length scales.

This review addresses the gap by unifying heat-transport physics across scales, modeling methods from atomistic to system level, cross-scale handoff workflows, validation and uncertainty, and design guidance for representative heterogeneous 3D stacks. Specifically, it compares atomistic interface models, mesoscopic transport formulations, continuum solvers, compact thermal and electrothermal models[12], reduced-order techniques, and emerging data-driven surrogates, while also connecting them to experimental validation pathways such as thermoreflectance imaging and thermoreflectance-based interface metrology [8], [13], [14]. The goal is not only to summarize prior work, but to clarify which model is appropriate at which scale and design stage, what information must be passed across scales, and where the major validation and uncertainty gaps still limit predictive thermal design.

## II. REPRESENTATIVE HETEROGENEOUS 3D STACKS AND THERMAL BOTTLENECKS

Before comparing modeling methodologies, it is necessary to define the representative thermal objects being modeled. To keep the discussion concrete and to avoid conflating distinct packaging regimes, this review uses three recurring stack archetypes: 1) logic + stacked memory systems, represented by logic paired with HBM-class memory; 2) logic-on-base-die / logic-on-logic stacks, represented by Foveros-style or other face-to-face vertically bonded logic tiers; and 3) chiplet/interposer-based heterogeneous packages, in which multiple active dies are laterally integrated on a passive or active interposer. These classes span the mainstream design space from package-level 2.5D integration to fine-pitch vertical integration, while exposing different balances among power density, vertical path length, lateral spreading, interface resistance, and cooling asymmetry [2], [6], [10].

Across all three archetypes, the thermal problem is defined not only by the total dissipated power but also by where heat is generated relative to the cooling boundary, which layers dominate the thermal resistance, and how strongly adjacent dies thermally couple to one another. In heterogeneous 3D systems, a neighboring die may act as a secondary heat source, a lateral heat spreader, or an additional thermal barrier depending on its material stack, thickness, and position in the package. Consequently, the same nominal power budget can yield markedly different temperature distributions across the three archetypes, even when the footprint and external cooling boundary are similar[6], [8], [13]. The purpose of this section is therefore not to catalog every available package option, but to establish the representative stack classes that will be reused in later sections when comparing model fidelity, computational cost, and validation requirements.

### A. LOGIC+HBM STACKS
Logic + HBM systems remain one of the most important heterogeneous integration archetypes for bandwidth-limited computing because they combine a thermally aggressive logic die with one or more vertically stacked memory towers (Fig.2). In current mainstream implementations, the compute die and HBM stacks are typically placed side by side on a silicon interposer, while more aggressive future variants bring the memory stack even closer to the logic tier. From a thermal standpoint, this architecture is characterized by an inherent asymmetry: the logic die usually dominates total power dissipation, whereas the memory stack often has tighter thermal headroom and is susceptible to both self-heating and imported heat from the adjacent logic die [15], [16].

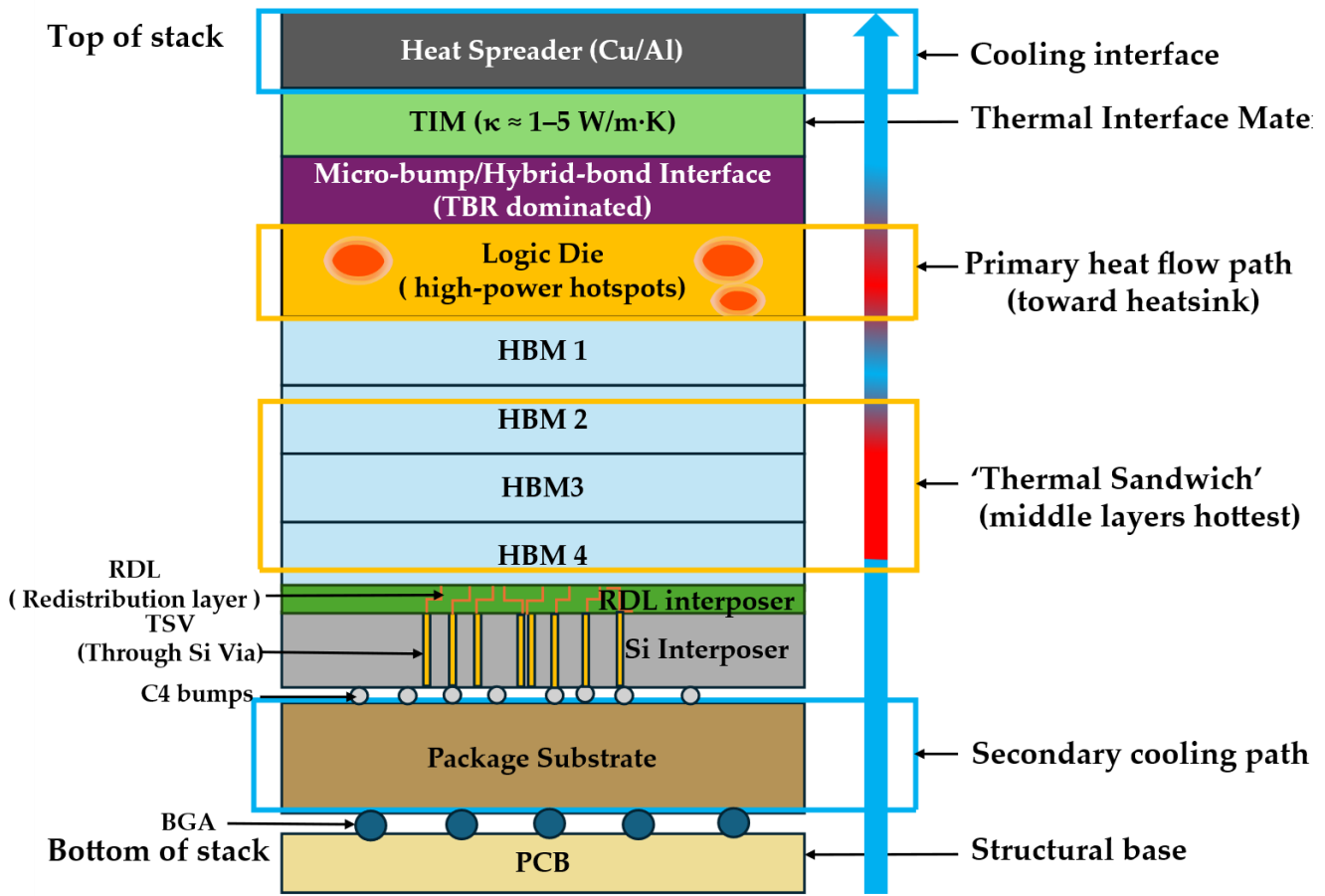

**FIGURE 2.** *Dominant heat-flow paths and vertical bottlenecks in a representative logic–HBM stack.*
High-power logic hotspots heat the overlying HBM stack; thermally resistive TIM and TBR-dominated micro-bump interfaces, combined with low-κ BEOL and HBM layers, cause middle dies to reach the highest junction temperatures, while secondary heat spreading occurs through the RDL/interposer and package substrate.

The first question is therefore where heat is generated. In this archetype, the dominant heat source is generally the compute die, particularly in accelerator-oriented workloads with high arithmetic intensity and dense on-die interconnect activity. However, stacked memory is not thermally passive. Heat is also generated in the HBM base die, memory PHY (Physical Layer) circuitry, and DRAM layers, and this self-heating is compounded by lateral and package-mediated coupling from the nearby logic die. Experimental and modeling work on stacked logic + WideIO and HBM-class structures shows that memory-side temperature cannot be predicted from memory power alone; rather, it depends on both internal dissipation and the thermal field imposed by the neighboring compute die and package boundary[15],[16]. In particular, Son *et al.* reported that, in an HBM-GPU module, thermal coupling from the GPU can become the dominant factor governing HBM thermal stress, which is why additional heat-spreading path between logic and memory are often explored [17].

The second question is what the dominant vertical bottlenecks are. For logic + HBM stacks, the key bottlenecks are the stacked memory dies themselves, the die-to-die bonding layers within the memory tower, and the final escape path from the memory/logic assembly to the lid, thermal interface material, and heat sink. The inner DRAM dies are especially disadvantaged because they are vertically buried and laterally confined, while the neighboring logic die can inject additional heat through the interposer and common heat-spreading structure. This makes the temperature field highly sensitive to hotspot alignment, stack height, and the detailed boundary conditions at the package level[6] [8], [16], [17].

The third question is which interfaces matter most. In this archetype, the thermally relevant interfaces include the die-to-die bonding layers within the memory stack, the HBM-to-interposer or HBM-to-logic attachment, and the package-side TIM/lid/heat-sink interfaces. Interface-dominated conduction becomes especially important because the memory stack contains repeated thin layers and buried boundaries, so the accumulated temperature rise depends not only on bulk silicon conduction but also on the resistance of multiple interfacial layers [8], [14]. What makes this archetype distinct thermally is therefore not simply that memory is stacked, but that a comparatively temperature-sensitive memory tower resides beside or above a much stronger heat source. For modeling, this implies that self-heating within the stack, cross-heating from the logic die, and package-level heat removal must all be represented consistently.

## B. LOGIC-ON-BASE-DIE / FOVEROS-STYLE STACKED LOGIC

Logic-on-base-die and logic-on-logic stacks differ from logic + HBM in one crucial respect: multiple vertically aligned tiers may each dissipate substantial power. Platforms such as Intel's Foveros demonstrate active-die stacking over a base die, while more recent fine-pitch hybrid-bonded logic-on-logic demonstrations push vertical interconnect density even further [5]. In contrast to logic + HBM systems, where the memory tiers are generally cooler than the main compute die, stacked-logic systems may place two or more thermally aggressive functional tiers within the same vertical column. As a result, hotspot superposition rather than simple thermal coupling becomes the dominant concern.

The first question again is where heat is generated. In stacked-logic systems, heat can be generated in the top logic tier, the base die, or both, depending on how compute, cache, SRAM macros, power delivery, and I/O functions are partitioned. This makes the thermal profile highly dependent on architectural decomposition. Early studies of die-stacked high-performance microprocessors showed that peak temperature increases when active layers are stacked farther from the heat sink and when high-power blocks are vertically aligned [18]. More recent sign-off-quality analysis of a commercial 3D microprocessor implemented with face-to-face bonding confirmed the same qualitative behavior: naively folding a high-performance CPU into multiple tiers increases thermal stress, whereas thermally aware functional partitioning can significantly reduce the penalty [19]. Thus, unlike logic + HBM, the dominant uncertainty in this archetype often lies not in whether the neighboring tier is hot, but in how strongly the hotspots of multiple active tiers overlap in the vertical direction.

The second question is what the dominant vertical bottlenecks are. In stacked-logic systems, the primary bottlenecks are the die-to-die bonding interface, the thin silicon tiers, and the BEOL/interconnect stack through which both signals and heat must pass. Because the upper tier is farther from the principal cooling boundary, its temperature rise depends strongly on the cumulative resistance of the lower active tier, interface layers, and package thermal path. This problem becomes more severe when several hot functional blocks are stacked with limited lateral offset. Accordingly, thermal behavior is highly sensitive to cross-tier floorplan correlation, tier assignment, and cooling direction[18], [19].

The third question is which interfaces matter most. Here, the die-to-die bonding interface and the thin interconnect/ILD stack are central, not just as parasitic details but as first-order determinants of thermal spreading. *Koroglu* and *Pop* showed that replacing conventional interlayer dielectrics with higher-thermal-conductivity insulators can materially reduce the spreading resistance of 3D logic stacks, emphasizing that BEOL anisotropy and thin-film material choice can no longer be neglected in vertically integrated logic [20]. What makes this archetype thermally distinct is therefore the coincidence of fine-pitch vertical interconnects, multiple buried active tiers, and strong hotspot superposition. For modeling, this means that power maps, BEOL anisotropy, interface resistance, and tier partitioning must be treated as coupled variables rather than as separable post-processing corrections.

## C. CHIPLET/INTERPOSER-BASED HETEROGENEOUS PACKAGES

Chiplet/interposer-based packages occupy the other end of the mainstream heterogeneous-integration spectrum. Instead of vertically overlapping multiple active tiers, they laterally place heterogeneous dies on a passive or active interposer. This generally relaxes the most severe penalty associated with buried vertical stacking, but it does not eliminate thermal bottlenecks. Rather, it shifts the dominant challenge from vertical hotspot superposition to lateral thermal crosstalk among nearby chiplets through the interposer, lid, and package heat-spreading structures[21], [22], [23], [24].

The first question is again where heat is generated. In this archetype, heat is generated primarily in the active chiplets rather than in the interposer itself. The resulting temperature field is therefore governed by the distribution of lateral heat sources, their spacing, their power imbalance, and the effectiveness of the common lid/heat-sink structure in redistributing heat. Zhou *et al.* showed that in a 2.5D TSV-interposer package, thermal behavior depends strongly on chiplet bonding approach, the lid and heat-sink geometry, the TIM selection, and thermal crosstalk between neighboring dies [21]. Importantly, they also reported that bumpless interconnects can be thermally advantageous relative to bump-based assembly, highlighting that even in a laterally integrated architecture, local interconnect technology materially affects heat flow [21].

## D. CROSS-ARCHETYPE SYNTHESIS AND MODELING IMPLICATIONS

Taken together, the three archetypes motivate different thermal abstractions. Logic + HBM systems require models that capture stack-internal interfaces, memory self-heating, and cross-heating from neighboring logic die under realistic package boundaries. Logic-on-base-die / logic-on-logic stacks require explicit treatment of vertically aligned hotspots, thin-layer anisotropy, and the thermal consequences of functional partitioning across active tiers. Chiplet/interposer packages require efficient but materially aware modeling of lateral crosstalk, lid/interposer spreading, and placement-dependent cooling behavior [8], [13], [19], [20]. These distinctions are the reason a single thermal abstraction is rarely sufficient across the full heterogeneous 3D design space. In the remainder of this paper, these three representative stacks are therefore used as recurring reference cases when assessing atomistic interface models, continuum solvers, compact thermal networks, reduced-order techniques, and data-driven surrogates.

## III. GOVERNING THERMAL PHYSICS AND CROSS-DOMAIN COUPLING

Heat transport in heterogeneous 3D ICs is not governed by a single abstraction that remains valid across all length scales. At the package and thick-die levels, diffusion-based heat spreading is often an adequate first approximation. In thin silicon tiers, BEOL/interconnect stacks, and bonded interfaces, however, size effects, interfacial resistance, and anisotropic effective conductivity become first-order terms rather than small corrections. Moreover, temperature does not remain confined to the thermal domain: it feeds back into electrical behavior through mobility, leakage, and resistive loss, and it couples into the mechanical domain through thermal expansion mismatch and thermal gradients. A predictive multiscale framework must therefore identify which physics dominates at which scale and preserves that physics when parameters are passed upward from atomistic or feature-scale models to continuum and compact models[13], [8], [24], [25].

At scales where local thermal equilibrium and diffusive transport remain valid, temperature evolution can be described by the transient anisotropic heat equation:

$$\rho C_p \frac{\partial T}{\partial t} = \nabla \cdot (K \nabla T) + Q \qquad (1)$$

where $\rho$ is the material density, $C_p$ is specific heat capacity of the material, $K$ is the thermal-conductivity tensor, and Q is volumetric heat source which includes contributions from joule heating, leakage currents, and other power dissipation mechanisms within active devices.

Equation (1) provides the continuum backbone for chip- and package-level modeling, but its coefficients and boundary conditions must be modified when the characteristic dimensions approach carrier mean free paths, when interfaces impose finite temperature jumps, or when temperature feeds back into electrical and mechanical behavior [13], [26].

### A. THIN FILM AND SIZE EFFECTS

Classical Fourier conduction implicitly assumes that the thermal conductivity of each material is a scale-independent bulk property. That assumption becomes progressively weaker in thin silicon films, BEOL dielectric layers, confined interconnect regions, and bonded structures whose characteristic dimensions approach the relevant phonon mean free paths[13], [26]. In such cases, boundary scattering and partial ballistic transport suppress heat flow relative to the bulk limit, so the conductivity used in a continuum model must be interpreted as an effective, geometry-dependent quantity rather than an intrinsic constant. This is particularly relevant in heterogenous 3D ICs because many thermally important layers are intentionally thin and vertically confined.

A useful way to express this scale dependence is through a suppression-function formulation for the effective thermal conductivity,

$$k_{eff}(L) = \int_0^\infty S\left(\frac{\Lambda}{L}\right) \frac{dk_{acc}}{d\Lambda} d\Lambda \qquad (2)$$

where $L$ is the characteristic length scale, $\Lambda$ is the phonon mean-free path, $S\left(\frac{\Lambda}{L}\right)$ is a transport-suppression function, and $k_{acc}$ is the cumulative thermal conductivity as a function of mean free path [11], [26]. Equation (2) makes clear that once $L$ becomes comparable to the dominant carrier mean free paths, the conductivity propagated into higher-level solvers is no longer a bulk value, but a reduced effective property derived from suppressed transport.

For multiscale modeling, the implication is straightforward: fine-scale calculations or measurements are often needed to extract corrected conductivities before applying die or package-scale diffusion solvers [13], [27]. If thin-film and size effects are omitted, bulk conductivities are implicitly assigned to layers that no longer behave as bulk media. The model then overestimates local heat spreading, underestimates hotspot severity, and obscures the real thermal penalty associated with buried thin tiers and confined conduction paths.

### B. THERMAL BOUNDARY RESISTANCE

Even when adjacent layers are individually well characterized, heat flow across their interface may be strongly impeded by imperfect transmission of vibrational energy. This interfacial penalty is described as thermal boundary resistance (TBR), or equivalently interfacial thermal conductance, and it becomes especially important in heterogeneous 3D ICs because such systems intentionally assemble many dissimilar materials through repeated bonded interfaces[8], [28]. The issue is not restricted to visibly poor contact; even atomically clean interfaces exhibit finite resistance because the energy carriers on the two sides do not couple perfectly [28]. More recent theoretical and simulation work also shows that simple elastic mismatch pictures are often insufficient once interface disorder, mode conversion, and inelastic scattering become important [8], [29].

At an interface, the heat flux can be written as

$$q'' = G(T_1 - T_2) = \frac{T_1 - T_2}{R_K} \qquad (3)$$

where $G$ is the interfacial thermal conductance and $R_K = \frac{1}{G}$ is the thermal boundary resistance. At the same time, flux continuity requires

$$-n \cdot K_1 \nabla T_1 = -n \cdot K_2 \nabla T_2 = q'' \qquad (4)$$

so that a finite temperature jump may exist even when the heat flux is continuous across the boundary [28], [29].

Equations (3) and (4) are especially important for vertically integrated stacks because bonded interfaces are not merely geometric separators; they are thermally active elements that may dominate the total vertical resistance when repeated across a short conduction path[8]. In micro-bumped stacks, hybrid-bonded assemblies, die-attach layers, and package-side TIM interfaces, the aggregate interface contribution can rival or exceed the bulk resistance of thin films or buried layers. If TBR is omitted, the model enforces an unphysical temperature continuity across interfaces, misattributes interface-controlled temperature rise to the bulk layers, and can substantially distort the apparent benefit of interface engineering or bonding-technology choices.

### C. MATERIAL ANISOTROPY IN BEOL AND INTERCONNECT STACKS

At larger scales, heat transport in 3D ICs is often represented using homogenized material properties. However, the BEOL and interconnect environment is not a uniform medium but a composite of Cu lines, vias, liners, barrier layers, low-k dielectrics, redistribution layers and bumps are arranged in strongly directional geometries. As a result, its effective thermal conductivity is anisotropic and layout dependent rather than isotropic and layer independent [30], [31]. Experimental work on advanced interconnect structures with composite low-k dielectrics showed that the effective thermal conductivity of embedded interconnect media can differ substantially from naïve bulk-material expectations, which is why interconnect heating and temperature rise cannot be inferred from standalone dielectric properties alone [31].

The directional nature of heat transport can be written as

$$q = -\mathbf{K}\nabla T, \quad \mathbf{K} = \begin{bmatrix} k_x & 0 & 0 \\ 0 & k_y & 0 \\ 0 & 0 & k_z \end{bmatrix} \quad (5)$$

where $k_x$, $k_y$, and $k_z$ are the effective conductivities along the principal directions. For many interconnect environments, $k_x$ and $k_y$ may be similar, while $k_z$ differs substantially because vertical heat flow must cross layered dielectric and interface structures[30], [32].

This anisotropy matters because local heat spreading and vertical escape are governed by different structural pathways within the BEOL stack. Cu-rich routing directions, via density, and dielectric selection affect lateral and cross-plane thermal transport unequally. Low-k and ultra-low-*k* materials, although beneficial for reducing interconnect capacitance, generally degrade thermal conductivity relative to more thermally conductive alternatives. Recent studies on 3D IC thermal design have further shown that replacing conventional low-thermal-conductivity insulators with higher-thermal-conductivity dielectrics can materially reduce thermal resistance, highlighting that BEOL material selection must be treated as a thermal as well as an electrical design variable[33]. At the chiplet and package levels, anisotropic homogenization is likewise needed to preserve the directional heat-spreading behavior of RDLs, TSV/interposer features, and composite attachment structures [24]. If anisotropy is ignored, the predicted hotspot shape becomes unrealistically symmetric, the relative importance of lateral versus vertical cooling is distorted, and the design value of BEOL, via, or dielectric modifications may be either understated or overstated.

### D. ELECTROTHERMAL COUPLING

Temperature in a heterogeneous 3D IC is not merely an output of power dissipation; it also perturbs the electrical behavior that generates that power. Device characteristics such as carrier mobility are temperature dependent[34], and circuit-level power components such as leakage current and resistive loss can increase with temperature [9]. This creates a positive electrothermal feedback loop (Fig. 3): higher temperature modifies electrical transport, altered electrical transport changes power dissipation, and the new power distribution further reshapes the thermal field. The coupling is especially important in vertically constrained stacks because buried tiers and interfacial resistance reduce thermal margin and make local temperature excursions harder to dissipate.

A compact continuum form of this coupling can be written as

$$\nabla \cdot (\sigma(T) \nabla V) = 0 \quad (6)$$

for the electrical potential $V$, where $\sigma(T)$ is the temperature-dependent electrical conductivity. The corresponding Joule-heating density is:

$$\dot{q}_J = \sigma(T)|\nabla V|^2 = \mathbf{J} \cdot \mathbf{E} \quad (7)$$

and the thermal problem becomes

$$\rho C_p \frac{\partial T}{\partial t} = \nabla \cdot (\mathbf{K} \nabla T) + \dot{q}_J + \dot{q}_{leak}(T) \quad (8)$$

where $\dot{q}_{\text{leak}}(T)$ represents temperature-dependent leakage-related heat generation[9], [34].

Equations (6)– (8) formalize electrothermal feedback by coupling the electrical solution to heat generation and the thermal solution back to temperature-dependent electrical behavior. The severity of this coupling depends on scale. At the device level, it governs how self-heating affects local carrier transport. At the block and die levels, it changes leakage maps, Joule-heating intensity, and workload-dependent hotspot evolution. At the stack level, it can determine whether a nominally acceptable thermal solution remains stable once temperature-dependent power is introduced. This is why electrothermal co-analysis is not merely a refinement for final signoff; it is often required to judge whether a high-power tier placement or cooling scheme is fundamentally viable. If electrothermal coupling is omitted, thermal analysis becomes a one-way post-processing step performed on a frozen power map. In power-dense heterogeneous 3D ICs, that can underpredict hotspot severity, shift the apparent hotspot location, and obscure instability margins associated with leakage escalation or thermal runaway [9].

### E. THERMOMECHANICAL IMPLICATIONS

The thermal field in a heterogeneous 3D IC also couples directly into mechanical stress through thermal expansion mismatch and nonuniform temperature gradients. This is especially pronounced in TSV-based and bonded 3D structures, where Cu, Si, dielectrics, underfills, and bonding layers expand differently during fabrication, assembly, and operation[25], [35]. As a result, the same temperature distributions targeted by thermal modeling can also drive stress concentrations near interfaces, vias, and localized hotspots. Those stresses are not only a packaging concern; they can perturb device mobility, influence timing, and degrade long-term reliability through cracking, delamination, fatigue, or stress-induced variation[36], [37].

A compact thermoelastic constitutive relation can be written as

$$\sigma = \mathbf{C}:(\varepsilon - \alpha(T - T_0)\mathbf{I}) \quad (9)$$

where $\boldsymbol{\sigma}$ is the stress tensor, $\mathbf{C}$ is the elastic stiffness tensor, $\boldsymbol{\varepsilon}$ is the strain tensor, $\alpha$ is the thermal-expansion coefficient, $T_0$ s a reference temperature, and $\mathbf{I}$ is the identity tensor. Mechanical equilibrium then requires

$$\nabla \cdot \sigma + b = 0 \quad (10)$$

where b is the body-force density.

Equations (9) and (10) show that the computed thermal field is also the forcing term for stress generation. A stack with an acceptable peak temperature may still be mechanically risky if it sustains strong vertical gradients or concentrates stress around TSVs, bonded interfaces, or localized hotspots. Conversely, some thermal-mitigation strategies may also reduce stress by redistributing heat more uniformly. If thermomechanical coupling is omitted, the model can declare a design thermally safe while missing reliability penalties induced by thermal gradients and material mismatch. In effect, the thermal model becomes disconnected from one of its most important design outcomes: whether the stack can survive fabrication, packaging, and field operation without stress-driven degradation.

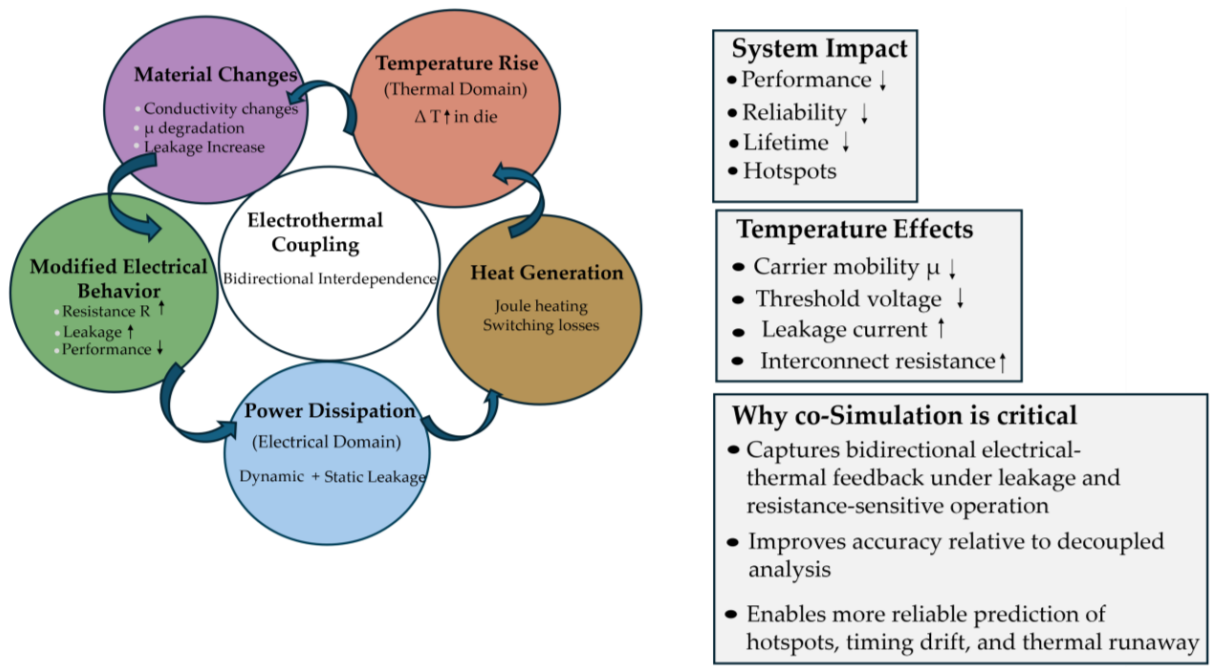

**FIGURE 3.** Electrothermal Coupling Feedback Loop and its implications for 3D IC analysis.
The figure summarizes how temperature-dependent electrical behavior can alter power dissipation and, in turn, reshape the thermal field, motivating coupled electrothermal simulation.

### F. CROSS-SCALE SYNTHESIS

Taken together, the governing physics suggests a practical hierarchy of dominant mechanisms. At the smallest scales, thin-film suppression and quasi-ballistic transport determine the effective conductivity that should be handed upward to larger-scale solvers[13], [26], [27]. At material boundaries, TBR determines whether vertical conduction is interface limited rather than bulk limited [8] [29]. At the BEOL and local interconnect scales, anisotropic effective properties govern directional heat spreading[33], [24], [30]. At the die and stack scales, electrothermal feedback determines whether power and temperature must be solved self-consistently[9], [34]. Finally, at the reliability and design-closure level, thermomechanical coupling determines whether the computed temperature field is compatible with mechanical integrity and timing robustness [25]–[38].

The modeling implication is straightforward: a thermal solver doesn't need to resolve every physics at every scale, but it must preserve the dominant mechanism before passing reduced parameters upward or downward. Omitting a mechanism at the scale where it dominates does not simply introduce a small numerical error; it changes the physical meaning of the model. That requirement is what distinguishes a physically consistent multiscale workflow from a collection of loosely connected thermal abstractions.

## IV. MULTISCALE THERMAL MODELING APPROACHES

No single thermal-modeling framework is uniformly optimal across the heterogeneous 3D design stack. The relevant choice depends on the physical scale of interest, the governing transport regime, the required outputs, and the downstream task to be supported. At the smallest scales, the central unknowns are phonon and electron properties, force constants, and interface-specific transport coefficients. At intermediate scales, the main objective is to capture partial ballisticity, nonlocal heat flow, and apparent material behavior in geometries whose dimensions are comparable to carrier mean free paths. At the die, package, and system levels, the main requirement shifts toward fast and robust prediction of temperature fields, thermal impedances, hotspot trajectories, and electrothermal interactions under realistic power maps and boundary conditions. The method landscape is therefore best understood as a hierarchy of modeling frameworks that resolve different state variables and export different handoff quantities to the next scale[9], [11], [12], [26], [28], [39], [40], [41].

In this section, the goal is not merely to enumerate methods, but to clarify what each framework is fundamentally best at, what information it produces, how that information is handed to the next scale, and where each method fails if it is used outside its natural operating regime. This viewpoint is especially important in heterogeneous 3D ICs, where the thermal problem is defined by repeated cross-scale handoffs: atomistic interface physics must be converted into interfacial conductance laws, thin-film transport into effective conductivity tensors, high-fidelity field solutions into compact RC networks, and expensive full-order solvers into reduced or learned surrogates for iterative design and runtime use.

### A. FIRST PRINCIPLES/DFT/DFPT

First-principles methods, typically based on density functional theory (DFT) and density functional perturbation theory (DFPT), provide the most fundamental description of thermal transport in crystalline materials. These methods can predict phonon dispersion, group velocity, mode heat capacity, anharmonic scattering rates, and intrinsic lattice thermal conductivity without empirical fitting, thereby offering a rigorous route to materials-property prediction [13], [37]. In the context of heterogeneous 3D ICs, their principal value lies in the evaluation of candidate bonding materials, dielectric layers, ultrathin films, and other emerging materials for which reliable thermal data are not yet available experimentally [37], [38].

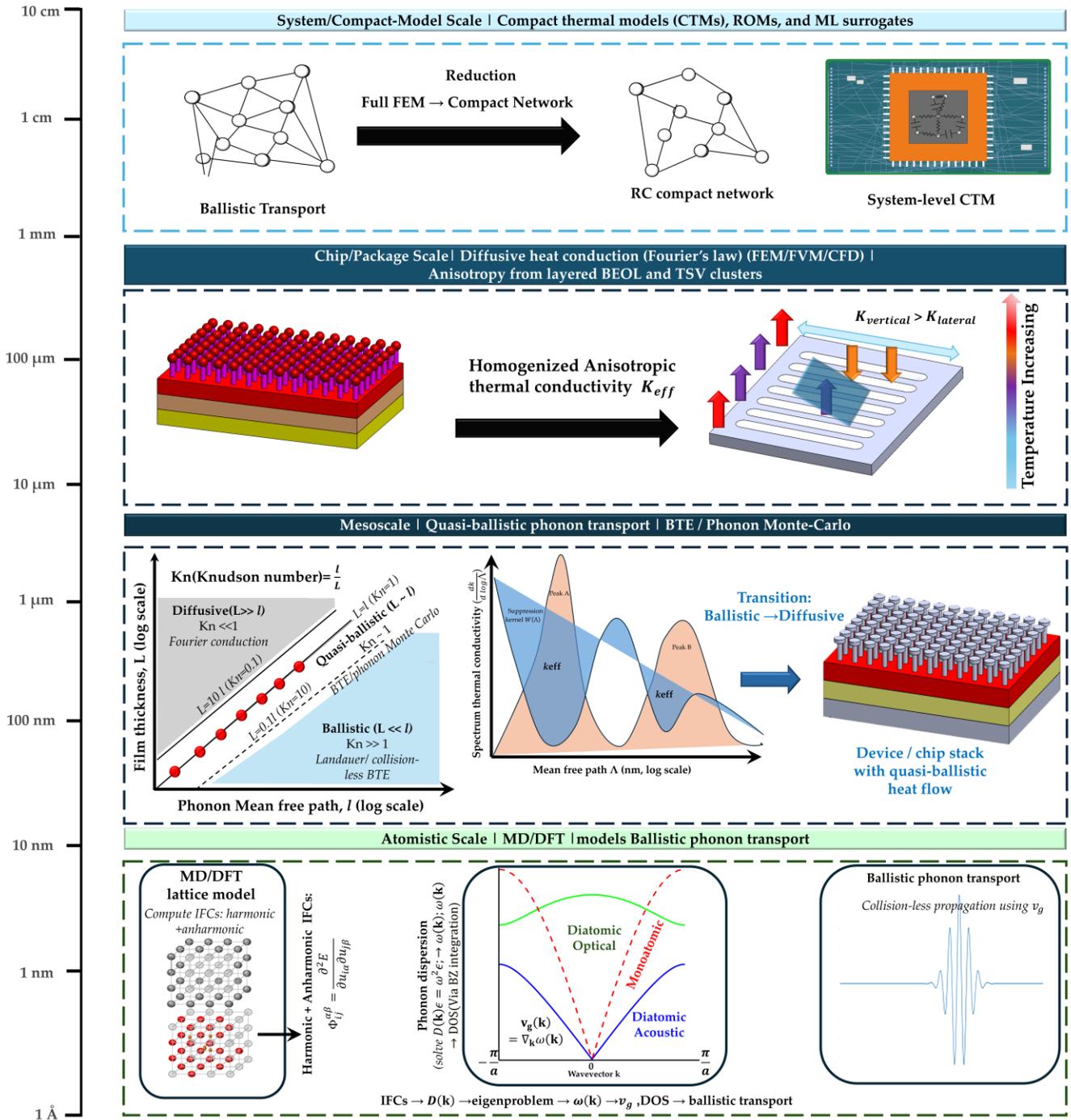

**FIGURE 4.** Multiscale thermal transport and modeling hierarchy for heterogeneous 3D ICs.
(a) Atomistic scale: ballistic phonon transport and phonon property extraction using MD/DFT models (1 nm–10 nm). (b) Quasi-ballistic phonon transport at device/system length scales, modeled by BTE and phonon Monte-Carlo (~100 nm–1 μm). (c) Chip/package scale: diffusive heat conduction governed by Fourier's law, solved via FEM/FVM/CFD, with anisotropy arising from layered BEOL stacks and TSV clusters (~1–10 μm to 1 cm). (d) System/EDA scale: compact thermal models (CTMs), reduced-order models (ROMs), and ML surrogates obtained by reducing detailed FEM models for efficient chip- and system-level analysis (~1 μm–10 cm).

However, the strength of first-principles approaches is also their main limitation. They are highly effective for idealized or near-ideal materials and interfaces, but much less suited to rough, defective, process-variable, or mechanically stressed interfaces of the kind encountered in real heterogeneous stacks. They also remain computationally prohibitive for direct simulation of practical device, die, or package structures. Accordingly, their role in 3D-IC thermal modeling is not direct full-stack prediction, but upstream parameter generation: they produce intrinsic transport quantities, spectral information, and reference material properties that can be passed on to atomistic, mesoscopic, or continuum models.

In practical heterogeneous 3D-IC workflows, first-principles methods are therefore best viewed as reference-level materials tools rather than deployable design-flow engines. Their maturity is high for intrinsic property prediction, but low for direct stack-level use. Their main handoff quantities are phonon properties, intrinsic conductivity, and electron–phonon coupling parameters. Their main failure mode is over-idealization of real interfaces and process variation. As a result, they are most useful when the design question concerns what a material could do in principle, not what a manufactured 3D stack will do in practice [13], [37], [38].

### B. MOLECULAR DYNAMICS

Molecular dynamics (MD) extends the analysis from idealized lattice properties to finite-temperature atomic motion and morphology-sensitive transport. Both equilibrium and nonequilibrium MD are widely used to evaluate thermal conductivity, interfacial thermal conductance, spectral mean free path behavior, and the effects of roughness, disorder, and local nonlinearity[42]. In heterogeneous 3D ICs, this makes MD especially valuable for studying Cu/dielectric boundaries, bonded interfaces, thin interlayers, nanostructured TIMs, and other local structures whose thermal response depends strongly on atomic configuration.

Compared with first-principles methods, MD is more naturally suited to imperfect interfaces and non-ideal structures. Compared with continuum approaches, it retains the microscopic mechanisms that govern interfacial scattering and morphology-sensitive transport. This makes it particularly useful for extracting TBR, size-dependent conductivity, and local interface laws that can later be passed upward into higher-level thermal models. Nevertheless, its predictive reach remains strongly limited by accessible length and time scales, and classical MD further depends on the fidelity of the chosen interatomic potential. Ab initio MD avoids some of those force-field limitations but is too expensive for routine use beyond very small systems.

For heterogeneous 3D-IC design, MD is best regarded as a local physics-extraction tool, not a whole-stack thermal solver. Its maturity is strong for interface studies and nanometer-scale parameter extraction but limited for direct design deployment. The main outputs handed upward are interfacial conductance, morphology-sensitive corrections, and size-dependent effective conductivity. Its main strength is explicit treatment of roughness, disorder, and anharmonicity. Its main failure mode is the combination of potential dependence and severe scale limits. In practice, MD is most valuable when the dominant unknown is local interface physics, not global package temperature.

### C. MESOSCOPIC TRANSPORT: BTE/PHONON MONTE CARLO/ QUASI-BALLISTIC CORRECTIONS

The mesoscopic regime lies between atomistic transport and continuum diffusion. Here, the characteristic feature sizes are too large for atomistic simulation yet too small, or too nonlocal, for uncorrected Fourier conduction to remain reliable. The phonon Boltzmann transport equation (BTE), phonon Monte Carlo techniques, and quasi-ballistic correction models are the principal tools in this regime [13], [26], [27], [43]. These methods preserve directional and modal transport information and can therefore capture ballistic-to-diffusive transition, boundary scattering, spectral nonequilibrium, and hotspot-scale suppression of effective conductivity.

In heterogeneous 3D ICs, these methods are most relevant in ultrathin silicon tiers, dense bonded regions, BEOL-scale structures, and local hotspot neighborhoods where the assumptions underlying continuum diffusion begin to weaken. Their greatest practical value is not full-stack package analysis, but bridge-modeling: they quantify how confinement, interface scattering, and hotspot size alter the transport parameters that should be supplied to continuum or compact models. Quasi-ballistic correction laws are especially attractive in this respect because they preserve the key physics while remaining easier to embed in higher-level workflows than full BTE solvers. NEGF-based phonon transport belongs to the same general family of non-Fourier methods, but in this context, it serves mainly as a highly specialized reference tool at very small scales rather than as a routine design-flow method [43], [44].

This class is physically important but operationally selective. For heterogeneous 3D ICs, mesoscopic transport methods are mature enough to support local refinement and constitutive correction, but they are not yet practical as routine whole-stack design engines. Their main handoff quantities are suppression functions, apparent conductivity, interface-scattering corrections, and hotspot-aware transport parameters. Their main strength is faithful treatment of nonlocal transport where continuum diffusion can be misleading. Their main failure mode is computational burden and coupling complexity in realistic 3D geometries. In practice, these methods are best used to answer the question where Fourier modeling breaks down, not to replace continuum modeling everywhere.

### D. CONTINUUM PDE MODELS: FEM/FVM/FDM/GREEN'S FUNCTION/SEMI-ANALYTICAL METHODS

Continuum PDE models remain the workhorse of practical thermal analysis for heterogeneous 3D ICs. At these scales, the governing equation is the anisotropic transient heat equation, solved with geometry-aware material properties, interfacial conditions, and package-side boundary constraints. Finite-difference and finite-volume methods are efficient for structured multilayer domains; finite-element methods offer superior flexibility for irregular geometries and detailed package structures; and Green's-function or semi-analytical approaches are particularly attractive when the geometry is layered and repeated thermal solves are required[45], [46].

This class is the main design-closure layer of the modeling hierarchy. It is the most natural framework for predicting full temperature fields, hotspot locations, thermal gradients, transient response, and design sensitivities at die, interposer, and package scales. It is also the principal source of reference solutions used to calibrate compact models, reduced-order models, and learned surrogates. For that reason, continuum solvers are not merely another entry in the method catalog; they are the backbone against which most deployment-oriented abstractions are built.

TABLE I
Comparison of Multiscale Electrothermal Modeling Frameworks Used in Device and System-Level Thermal Analysis

| Modeling framework | Typical Scale | Core physics/ governing description | Electrothermal coupling fidelity | Key advantages | Main limitations |
|---|---|---|---|---|---|
| First-principles DFT/DFPT | Atomic (Å) to few-nm | Electronic structure, phonon dispersion, anharmonic force constants, electron–phonon transport | Explicit microscopic | Parameter-free; captures quantum confinement, interface, intrinsic material properties | Small domain and time scale; impractical for full 3D device simulation |
| Molecular dynamics (EMD/NEMD, ab initio MD for very small systems) | Few nm to ~100 nm | Atomistic heat transport with defects, interfaces, roughness, anharmonic scattering | Weak to moderate unless extended with electron models | Captures defects, nonlinearity, and interfacial thermal resistance | Accuracy depends on interatomic potentials; limited quantum carrier treatment; short timescales |
| NEGF (nonequilibrium Green's function) | Sub-10 nm to tens of nm | Quantum electron/phonon transport, tunneling, ballistic conduction, interface transmission | Explicit microscopic | Ideal for quantum-confined nanoscale devices | Poor scalability to large 3D domains; complex contact/self-energy treatment |
| Boltzmann transport equation / electrothermal Monte Carlo | Tens of nm to μm | Carrier/phonon kinetic transport with scattering and nonequilibrium dynamics | Strong self-consistent | Captures ballistic-to-diffusive transition, and hotspot formation | Computationally expensive in 3D; difficult boundary treatment; Monte Carlo noise |
| Continuum electrothermal PDEs (DD* / ET* / hydrodynamic + heat equation; FEM/FVM) | Sub-μm to cm | Poisson + carrier continuity + energy balance + Fourier heat conduction | Strong at device/package level | Mature TCAD/Multiphysics workflow; supports realistic device geometries | Reduced accuracy in ballistic or quantum-regimes |
| Hybrid multiscale / domain-decomposed models | nm to cm | Quantum/BTE near hotspots coupled with continuum heat equation elsewhere | Explicit cross-scale | Balances fidelity and computational tractability | Complex coupling and interface consistency |
| Compact thermal / electrothermal circuit models (CTM, Foster/Cauer, SPICE-compatible RC networks) | Device/ package/ circuit scale | Lumped thermal impedances and temperature-dependent device models | Circuit-level bidirectional | Extremely fast; suitable for circuit simulation and reliability analysis | Limited spatial detail: calibration often required |
| Reduced-order electrothermal models (POD/Krylov/parametric MOR) | Derived from full-order FE/ FVM models | Projection-based reduction of electrothermal state equations | Strong, if built from coupled full-order models | Near-FE accuracy with major speed-up | Re-reduction needed when geometry or parameters change |
| ML / PINN / surrogate electrothermal models | Variable | Learned mappings or physics-informed PDE solvers | Data-driven or physics-informed | Real-time prediction, rapid design exploration, digital twins, inverse problems | Generalization risk; training data and uncertainty challenges |

*Drift-Diffusion
*Energy Transport

At the same time, their apparent maturity can be misleading if the lower-scale inputs are weak. A continuum solver may converge numerically while still producing a physically incomplete answer if its thermal conductivity tensors, interfacial resistances, or boundary conditions are poorly calibrated. Thus, the main limitation of continuum modeling in heterogeneous 3D ICs is not the solution machinery itself, but the quality of the effective properties and interface laws supplied to it. In practical terms, continuum PDE models are currently the most mature and most deployable framework for full-stack thermal prediction, but their predictive value depends strongly on the fidelity of the multiscale handoff beneath them.

*E. COMPACT THERMAL AND ELECTROTHERMAL CIRCUIT MODELS*

Compact thermal models replace a continuous temperature field with an equivalent network of thermal resistances, capacitances, or macromodel states. Their greatest strength is computational efficiency: they are fast enough to support floorplanning, architecture exploration, thermal-aware scheduling, electrothermal circuit simulation, and runtime control [12], [46], [47]. In heterogeneous 3D ICs, where the same thermal configuration may need to be evaluated thousands of times under changing power maps or workloads, this speed advantage is often decisive.

Compact models are therefore the most natural bridge between thermal physics and system design. They are readily embedded into SPICE-like electrothermal loops, architectural simulators, and thermal control frameworks, and they allow bidirectional electrical–thermal interaction to be evaluated far more rapidly than full-order PDE solutions. Their utility is especially high when a reliable full-order or experimentally calibrated reference model already exists. Under those conditions, compact models provide the most practical representation for design exploration and runtime mitigation. Their limitation is that they gain speed by sacrificing spatial fidelity. Sharp hotspots, anisotropic spreading, and interface-dominated drops can be smeared out if the topology is too coarse or the calibration domain is too narrow. As a result, compact models are highly mature as deployment models, but much less trustworthy when used as standalone predictive solvers in new geometries, new bonding schemes, or poorly calibrated stacks. In short, compact thermal models are among the most useful frameworks in practical 3D-IC flows, but only when treated as calibrated surrogates of richer physics, not as substitutes for that physics.

### F. REDUCED-ORDER AND MULTI-FIDELITY MODELS

Reduced-order modeling (ROM) occupies the space between full-order continuum simulation and compact deployment models. The goal is to preserve the dominant thermal response of a large system while dramatically reducing the number of degrees of freedom, often through projection-based techniques, parametric model-order reduction, or other state-space compression strategies [48]. In heterogeneous 3D ICs, ROM is particularly attractive because many high-value tasks design-space exploration, control, inverse calibration, and uncertainty propagation require repeated solutions of closely related thermal problems.

Multi-fidelity approaches extend this idea by assigning different accuracy levels to different regions or stages of the workflow. For example, a coarse stack-scale model may be used globally while local refinement is reserved for hotspot regions, bonded interfaces, or thermally sensitive tiers. This strategy is especially effective in chiplet/interposer systems and stacked logic designs, where only a limited subset of the geometry may require very high spatial fidelity at a given time. In that sense, ROM and multi-fidelity modeling are not separate physical theories, but computational strategies for preserving the useful parts of a high-fidelity solution while avoiding its full cost.

These methods are increasingly practical, but their robustness remains conditional. They are most mature when the geometry family, parameter domain, and operating range are relatively stable. Their main handoff quantities are reduced bases, parametric response surfaces, and compact macromodel states. Their main strength is near-high-fidelity accuracy with substantial runtime reduction. Their main failure mode is loss of validity outside the reduction manifold, especially when topology changes, nonlinearities intensify, or new stack features enter the design. Accordingly, reduced-order methods are best treated as bounded-domain accelerators, not universally reusable thermal abstractions [22], [32], [47], [48].

### G. ML/PINN/SURROGATE MODELS

Machine-learning-based thermal modeling is best viewed as a family of accelerators and inverse models rather than a single method class. At least three subfamilies are now relevant. First, conventional supervised surrogates map measured or simulated features to temperatures or hotspot metrics for fast runtime use [49]. Second, physics-informed neural networks (PINNs) solve governing equations directly by minimizing PDE or BTE residuals, thereby reducing dependence on labeled data [50]. Third, operator-learning models learn mappings from entire input functions such as power maps, boundary conditions, or heat-transfer coefficients to output temperature fields, which is particularly attractive for repeated 3D-IC thermal solves across a family of related designs [51], [52], [41], [40].

For heterogeneous 3D ICs, this family is especially promising because conventional design loops repeatedly solve closely related thermal problems. Operator-learning approaches such as DeepOHeat and later 3D-IC-focused operator models target exactly this use case, while recent transient-prediction models focus on fast temporal response estimation for stacked designs. At the lower scales, PINN-BTE and differentiable BTE solvers suggest a different opportunity: they can approximate or accelerate mesoscopic transport solvers while retaining explicit governing equations and parameter dependencies. Despite these advances, ML surrogates should not be treated as unconditional replacements for physics solvers. Their reliability depends on training-domain coverage, physical admissibility, treatment of boundary-condition variation, and the presence of trustworthy error assessment. In other words, ML provides the largest practical benefit when it is embedded in a multiscale workflow with strong reference data, physics constraints, and validation loops, rather than when it is used as an unconstrained black box regressor (Fig.4).

This class remains the least mature in terms of standardization and trustworthiness. Its main outputs are fast field predictions, inverse parameter estimates, and runtime-ready predictors. Its main strength is dramatic speed after training and natural support for repeated design-space queries. Its main failure mode is silent loss of fidelity under distribution shift, incomplete physical admissibility, and uncertain extrapolation beyond the training manifold. Therefore, ML/PINN/surrogate models are best used today as accelerators and digital-twin components, not as standalone signoff tools.

### H. CROSS FRAMEWORK SYNTHESIS

Taken together, these frameworks are not equally mature for the same task. Continuum PDE solvers and compact thermal models are currently the most mature and most deployable for stack- and package-level design workflows, especially when supported by calibrated material and interface parameters [12], [43]–[47], [58]. Reduced order and multi-fidelity methods are increasingly practical for optimization, control, and iterative exploration, but remain domain dependent [48]. By contrast, first-principles methods, molecular dynamics, and mesoscopic transport models are most valuable as upstream physics and parameter-extraction tools rather than routine full-stack design engines [37]–[42], [52]–[54]. Finally, ML/PINN/surrogate approaches are promising but comparatively less standardized, with their most credible current role lying in acceleration, inverse modeling, and runtime prediction on top of validated physics-based workflows [49]–[55].

This distinction is important because it clarifies what is genuinely mature, what is conditionally deployable, and what remains primarily research oriented. For heterogeneous 3D ICs, the most reliable modeling strategy is not to seek a single universal solver, but to combine upstream physics generation, continuum design closure, and deployment-oriented reduction or learning in a calibrated and scale-consistent workflow. That workflow perspective provides the bridge to the next section, which focuses explicitly on cross-scale model handoff and calibration.

## V. TAXONOMY OF CROSS-SCALE MODELING WORKFLOWS

In section IV, we reviewed the main thermal-modeling frameworks across atomistic, mesoscopic, continuum, compact, reduced-order, and learned representations. In practical heterogeneous 3D IC design, however, these models are rarely used in isolation. Instead, thermal analysis is implemented as a cross-scale workflow, in which different solvers exchange material parameters, interfacial properties, fields, reduced states, or measurement-informed corrections across scales and, in some cases, across physical domains. The central question is therefore not only which solver is most accurate at a given scale, but also how information should be transferred, coupled, calibrated, and reduced so that the dominant physics is preserved while the overall workflow remains computationally tractable.

This distinction is especially important for heterogeneous 3D ICs because the relevant thermal bottlenecks are inherently distributed across scales. Thin-film suppression and interface resistance originate at nanometer to submicron scales, anisotropic heat spreading emerges from BEOL/interconnect structure, and electrothermal and package-level thermal interactions determine the final temperature field at the die and system scales. As a result, full-stack prediction depends as much on workflow design as on the fidelity of any individual solver. A useful conceptual basis is provided by prior multiscale transport literature, which distinguishes decoupled schemes, where information is passed upward through reduced parameters, from coupled schemes, where models exchange state information during the solve itself. In the heterogeneous 3D IC setting, that distinction naturally expands into a broader taxonomy of bottom-up parameter-passing workflows, concurrent local–global coupling workflows, measurement-calibrated inverse workflows, full-order-to-compact deployment workflows, and surrogate/runtime digital-twin workflows [13], [12], [46], [47]–[53].

### A. DECOUPLED BOTTOM-UP PARAMETER-PASSING WORKFLOWS

The most common workflow in heterogeneous 3D IC thermal analysis is the decoupled bottom-up scheme, in which lower-scale models are first used to extract quantities that are then supplied to a higher-scale solver. Typical examples include obtaining interfacial thermal conductance from atomistic simulation or thermoreflectance-based characterization, deriving thin-film or size-corrected effective conductivity for confined layers, or extracting anisotropic equivalent conductivity tensors for BEOL/interconnect structures before solving the die- or package-scale heat equation. In this workflow, the scales are not solved simultaneously; rather, the lower-scale physics is compressed into a small set of handoff quantities such as $k_{eff}(L)$, $G_{int}$, $R_K$, or $K_{eff}$, which are then used by continuum or compact models[13], [26], [27], [54].

This approach is particularly appropriate when the global stack geometry is known but key constitutive parameters are not. In logic–HBM systems, for example, bottom-up handoff may be used to propagate interface-specific TBR or thin-film corrections into a package-scale FEM/FVM model. In logic-on-logic stacks, the same strategy may be used to derive anisotropic effective properties for fine-pitch interconnect regions or bonded tiers before compacting the result into a reduced thermal network. In chiplet/interposer systems, feature-scale anisotropy or interface corrections may be passed upward into lateral crosstalk solvers or placement-aware package models. The attraction of this workflow is its modularity: each model is used where it is most physically natural, while the higher-level solver remains fast enough for stack-level evaluation.

The main limitation is that the handoff itself can become the dominant error source. If a strongly context-dependent interfacial or quasi-ballistic phenomenon is collapsed into a single scalar parameter that is later used far outside the conditions under which it was extracted, the upward transfer hides the very scale dependence the fine-scale model was meant to capture. Thus, the decoupled workflow is attractive for design iteration, but it depends critically on choosing handoff quantities that remain meaningful across the intended operating range.

### B. CONCURRENT LOCAL-GLOBAL COUPLED WORKFLOWS

A second workflow class is the concurrent or local–global coupled scheme, in which different scales exchange information during the solution process rather than only through one-time parameter transfer. In the general multiscale transport literature, this class is often described using concepts such as handshake and padding regions, where different physical descriptions overlap and continuity constraints are imposed across the coupling interface [11]. For thermal analysis, the practical goal is to preserve local non-Fourier or interface-sensitive physics only where it matters, while allowing the rest of the stack to remain in a computationally cheaper continuum representation.

For heterogeneous 3D ICs, concurrent workflows are most useful in hotspot or bottleneck zoom-in problems. A local bonded interface, ultrathin active tier, or sub-micron heat source may require a higher-fidelity mesoscopic treatment, while the surrounding die, interposer, and package can still be modeled diffusively. Representative examples include ballistic–diffusive domain decomposition and hybrid phonon Monte Carlo–diffusion schemes, in which a high-fidelity solver is applied only in the confined region and is coupled to a diffusion solver elsewhere through temperature and flux matching [54],[55]. The same logic can be adapted to 3D IC thermal design when hotspot-scale transport near hybrid-bonded tiers or interface-sensitive structures must be resolved without forcing the entire package into the cost regime of the most expensive solver.

The strength of this workflow is localized fidelity without full-stack atomistic or mesoscopic cost. Its main challenge is coupling consistency. If temperature, heat flux, or spectral energy exchange is not transferred correctly across the interface between the two model domains, the workflow can introduce numerical artifacts or artificial resistance at the model boundary. Consequently, concurrent workflows are physically appealing for the most demanding 3D-IC bottlenecks, but they are harder to deploy robustly than decoupled schemes in routine EDA flows.

## C. MEASUREMENT-CALIBRATED AND INVERSE WORKFLOWS

A third class is the measurement-calibrated and inverse workflow, in which experiments are not used only for end-of-paper validation, but are explicitly embedded in the multiscale chain. This is especially important in heterogeneous 3D ICs because many of the most important thermal parameters: bonded-interface resistance, effective conductivity of composite layers, local boundary conditions, and contact-related spreading penalties are strongly process dependent. As a result, nominal database values are often insufficient for predictive stack modeling.

In this workflow, measurements from thermoreflectance imaging, TDTR/FDTR, buried-interface thermometry, test structures, or on-die sensors are used to infer the latent parameters of the model. Instead of assuming that $k_{eff}(L)$, $G_{\text{int}}$, $h$, or contact resistances are known as a priori, the workflow solves an inverse problem: thermal data are compared against forward simulations, and the unknown parameters are adjusted until the discrepancy is minimized [14],[56],[57]. For heterogeneous 3D stacks, this is particularly valuable when buried interfaces cannot be accessed directly, or when the target quantity is not a bulk property but an effective parameter of a multilayer bonded region.

This workflow is central to credibility because it directly connects multiscale simulation to fabricated hardware. It is also the natural bridge into your later section on validation and uncertainty. Its main weakness is non-uniqueness: different parameter combinations may fit the same limited data equally well, especially when observability is poor. For that reason, the best inverse workflows combine multiple measurement modalities and calibrate not just one parameter at a time, but the full chain of interface, material, and boundary-condition assumptions.

## D. FULL-ORDER-TO-COMPACT AND REDUCED-ORDER DEPLOYMENT WORKFLOWS

A fourth class begins from a full-order field solution and then transforms that solution into a cheaper representation for repeated design use. This includes the extraction of compact RC thermal networks, reduced bases, parametric macromodels, and other deployment-ready abstractions from FEM/FVM or other continuum solvers. In practice, this workflow is the main route by which high-fidelity thermal simulation becomes usable in floorplanning, electrothermal co-simulation, optimization, and dynamic management.

For heterogeneous 3D ICs, this workflow is often the practical center of the design flow. A detailed field solver is first used to resolve the temperature distribution of a logic–HBM module, stacked-logic assembly, or chiplet package under representative power maps and package boundary conditions. That field solution is then reduced into a compact model that can be embedded in design-space exploration loops, architectural studies, or control-oriented analysis. HotSpot, 3D-ICE, MTA, and later reduced-order approaches all support this general pattern, although they differ in geometric fidelity, cooling support, and nonlinear capability[12], [46], [47], [48], [58].

The main advantage is that the workflow preserves a traceable link to full-order physics while enabling many repeated evaluations. Its main failure mode is domain brittleness. A reduced model built for one stack topology, one cooling configuration, or one range of power-map variation may fail once the design moves outside that calibration manifold. Accordingly, the quality of the workflow depends not just on the reduction method, but on whether the retained basis or compact topology spans the design space that the model will encounter.

## E. SURROGATE AND RUNTIME DIGITAL-TWIN WORKFLOWS

The fifth class is the surrogate or runtime digital-twin workflow, in which full-order simulations, reduced models, and measurements are combined to train fast predictive models for inner-loop optimization or online use. In heterogeneous 3D ICs, this can take several forms: regression models for hotspot prediction, operator-learning models that map power maps and boundary conditions directly to temperature fields, PINN-based inverse or forward solvers, and hybrid physics-plus-ML predictors embedded in runtime monitoring loops [49]–[55].

What distinguishes this workflow from reduced-order deployment is that the final mapping is not necessarily derived by explicit projection of the governing equations. Instead, the mapping is learned from data often generated by a continuum or compact solver and then augmented by measurements or physics constraints. This is particularly attractive in 3D IC thermal design because many solvers are structurally similar: the geometry class is fixed, while the dominant variations arise from changing power maps, workloads, boundary conditions, or design parameters. Operator-learning approaches such as DeepOHeat and recent transient prediction models directly target that setting by amortizing the cost of repeated PDE solves across large design families [51], [52].

The strength of this workflow is very low online cost after offline training. Its main failure mode is trustworthiness under distribution shift. If workload statistics, cooling conditions, aging, or fabrication variation move outside the domain represented in the training set, the surrogate may lose physical fidelity without obvious warning. For that reason, surrogate workflows are most defensible when they are built on top of validated full-order or reduced-order models and periodically updated using measurement feedback.

## F. CROSS-WORKFLOW SYNTHESIS

Taken together, these workflow classes differ primarily in how information moves across scales. Decoupled schemes pass reduced parameters upward. Concurrent schemes exchange state variables during the solve. Measurement-calibrated schemes feed experimental information back into the model chain. Full-order-to-compact schemes compress expensive field solutions into reusable macromodels. Surrogate workflows learn fast mappings from those calibrated models and data.

No single workflow is universally superior. For material and interface studies, bottom-up parameter-passing is often the right choice. For localized quasi-ballistic or interface-sensitive hotspots, concurrent local–global coupling is more appropriate. For signoff-quality predictive modeling, measurement-calibrated inverse workflows are essential. For EDA integration, floorplanning, and control, compact and reduced-order deployment workflows are usually the practical center. For runtime prediction and digital twins, surrogate workflows

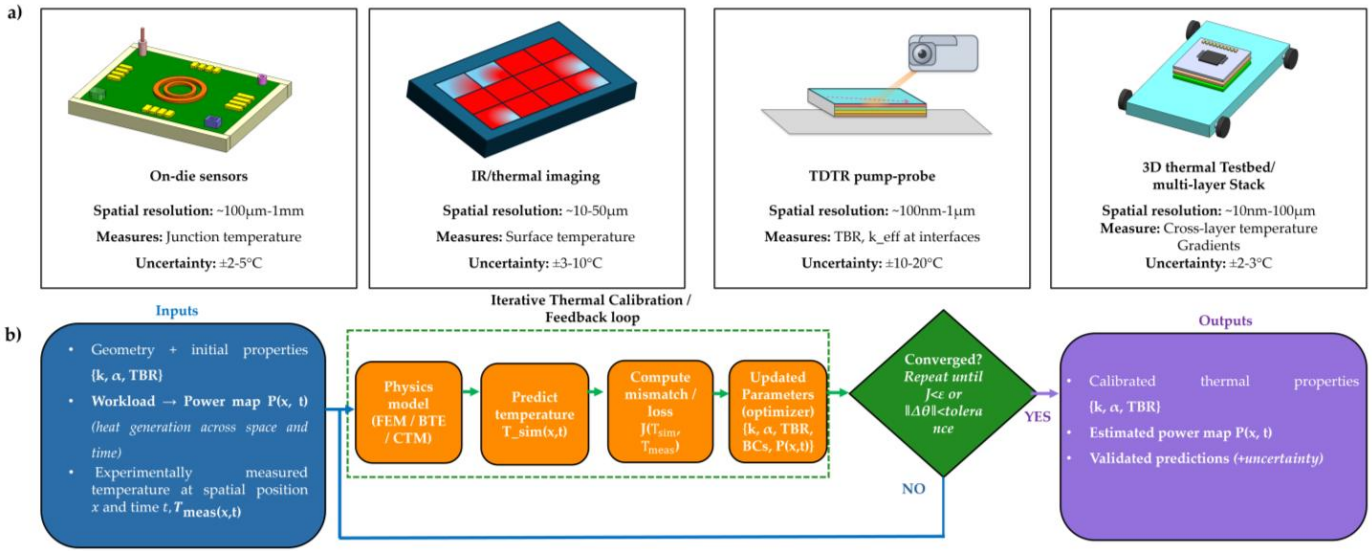

**FIGURE 5.** Experimental validation and iterative model calibration workflow for 3D IC thermal analysis
(a) Experimental thermal measurement techniques for heterogeneous 3D ICs: on-die sensors (junction temperature sensing with ~100 µm–1 mm spatial resolution and ±2–5 °C uncertainty), IR/thermal imaging (~10–50 µm, surface temperature, ±3–10 °C), TDTR pump–probe (~100 nm–1 µm, interfacial TBR and effective thermal conductivity, ±10–20 %), and 3D thermal testbeds / multi-layer stacks (~10 nm–100 µm, cross-layer temperature gradients, ±2–3 °C). (b) Model validation and learning loop linking physics-based FEM/BTE/CTM models to measurements: models generate spatial–temporal temperature predictions that are compared with measurements under realistic workloads (SPEC/MLPerf traces); discrepancies drive a feedback loop that updates material properties (TBR, κ, α), refines thermal models and boundary conditions, trains ML surrogates, and iteratively improves power maps used for chip- and system level electro-thermal analysis.

become attractive once the offline model chain has already been validated.

The key implication for heterogeneous 3D ICs is therefore that the hard problem is not only selecting a solver but designing a physically consistent workflow in which the handoff quantities, calibration strategy, and reduction layers remain aligned with the dominant physics at each scale. That workflow perspective is what converts a collection of modeling methods into a predictive multiscale design methodology.

## VI. EXPERIMENTAL VALIDATION, CALIBRATION, AND UNCERTAINTY

A multiscale thermal model is only as credible as the validation chain that supports it. In heterogeneous 3D ICs, this issue is particularly acute because many of the most important quantities in the model are not directly observable at the scale where they matter most. Junction temperature may be sampled only sparsely by embedded sensors; buried-interface conductance is usually inferred through model-based thermoreflectance metrology; and package-level temperature fields are often reconstructed from a combination of optical imaging, test vehicles, and reduced-order models. As a result, experimental validation is not a final consistency check added after simulation, but a central part of the modeling workflow that determines which parameters are trustworthy, which are only weakly identifiable, and what confidence can be assigned to the final temperature prediction [11] [59]-[60].

This section therefore treats validation, calibration, and uncertainty as a unified hierarchy. On-die sensors provide in situ temperature observations during operation. Optical thermography methods such as infrared imaging, thermoreflectance, and Raman spectroscopy provide spatially resolved thermal fields near accessible surfaces. TDTR and FDTR enable extraction of interface conductance and thin-film properties, including in buried multilayers through model-based inversion. Dedicated thermal test vehicles isolate specific thermal pathways under controlled heating conditions. Workload-based validation then connects these measurements to realistic power maps and thermal transients. Finally, inverse calibration and uncertainty propagation determine how measurement noise, model-form error, and parameter non-uniqueness affect chip- and package-level thermal predictions [59]–[75].

### A. ON-DIE SENSORS

On-die thermal sensors provide the most direct route to in situ temperature observation during circuit operation. In CMOS platforms, the underlying sensing element may be implemented using parasitic BJTs, MOS-based structures, ring oscillators, or digital thermal-sensor architectures, with the choice governed by area, accuracy, supply dependence, and the intended integration density. Reviews of CMOS temperature sensing emphasize that such sensors are attractive because they can be embedded directly into the chip and sampled during normal operation, while architecture-level thermal-monitoring work shows that they are essential for runtime thermal management and hotspot tracking in modern processors sensors[61][62],[63].

In the context of heterogeneous 3D ICs, on-die sensors directly measure only local electrical proxies of temperature at discrete locations. What they enable the modeler to infer is much broader: steady and transient hotspot behavior, thermal time constants, workload-dependent temperature trajectories, and the fidelity of compact thermal or electrothermal models under real operating conditions. Their main advantages are temporal accessibility and direct compatibility with runtime or workload-driven validation. Their main limitations are sparse spatial coverage, calibration drift, placement dependence, and the fact that buried tiers or neighboring dies may remain only indirectly observable. Consequently, on-die sensing is most powerful when used not as a standalone truth source, but as one layer in a wider validation stack that also includes optical metrology and calibrated full-field simulation.

### B. INFRARED IMAGING, THERMOREFLECTANCE, AND RAMAN THERMOGRAPHY

Wide-field optical techniques provide spatial information that on-die sensors cannot. Infrared thermography offers straightforward surface-temperature mapping over large fields of view, which makes it useful for package-level screening and qualitative hotspot localization. Its limitations, however, are equally important: spatial resolution is modest relative to microelectronic hotspot dimensions, emissivity must be handled carefully, and line-of-sight access becomes problematic once surfaces are obscured by passivation, lids, or package structures. For those reasons, IR imaging is best viewed as a useful but relatively coarse measurement layer rather than a universal validation tool for buried or fine-featured 3D stacks[59].

Thermoreflectance imaging offers a significantly finer optical route by converting small reflectivity changes into temperature maps. Recent reviews describe it as a high-resolution technique for device-temperature measurement and thermal-property characterization, with relevance to micro- and submicron hotspots when an appropriate thermoreflectance coefficient is established for the surface or transducer layer[59]. Raman thermography provides a complementary capability. Because Raman peak position and linewidth depend on local temperature, Raman-based methods can deliver submicron spatial resolution and, in some configurations, very fast temporal response, while also being useful for validating device-scale thermal models and extracting thermophysical properties of relevant materials [64].

### C. TDTR AND INTERFACE METROLOGY

If optical thermography provides spatial temperature fields, TDTR and FDTR provide access to the thermal parameters that generate those fields. Modern TDTR/FDTR methodology can be used to infer cross-plane thermal conductivity, volumetric heat capacity, anisotropic conductivity components, and interfacial thermal conductance in thin films, multilayers, and heterogeneous interfaces through model-based fitting of pump–probe signals[65]. More recent work has extended this capability to difficult buried-interface problems, including inverse extraction of thermal properties from FDTR data and improved sensitivity to buried thermal resistance through microscale confinement strategies micro-bumps have quantified how bump pitch and underfill affect interface thermal resistance[56], [66].

This is highly relevant to heterogeneous 3D ICs because many of the most important bottlenecks are hidden inside bonded stacks rather than exposed to the top surface. Die-to-die bonding layers, underfill-adjacent regions, thin composite films, and buried dielectric interfaces often dominate the vertical resistance chain but are inaccessible to direct line-of-sight thermography. TDTR and FDTR address that gap by turning the validation problem into a constrained inverse problem: the optical signal is measured, a layered thermal model is solved, and the unknown interface or thin-film parameters are adjusted until the response matches the data [66]–[69]. The output is therefore rarely a "direct measurement" in the naive sense; it is a statistically conditioned estimate of thermal properties under a chosen forward model.

The strength of TDTR/FDTR in this review context is that these methods directly support parameter identification for multiscale workflows. The main limitation is identifiability. In multilayer stacks, different combinations of conductivity, heat capacity, thickness, and interface conductance can produce similar optical responses unless the experiment is designed to separate their sensitivities. That is why recent inversion and uncertainty-analysis work has become so important: it clarifies which quantities can be inferred robustly, under what measurement configurations, and with what confidence bounds[56], [67].

### D. DEDICATED TEST VEHICLES

Dedicated thermal test vehicles remain one of the most valuable validation instruments for 3D integration because they isolate thermal pathways under controlled, repeatable conditions. In the 3D-IC context, test vehicles typically integrate distributed heaters (Fig. 5) and embedded temperature sensors so that vertical and lateral heat spreading can be excited intentionally and compared against simulation. IMEC's thermal test-vehicle work is particularly relevant here: the ESTC 2010 study introduced a dedicated 3D stacked thermal test vehicle for validating hotspot-dissipation modeling, and the subsequent Microelectronics Journal paper demonstrated fine-grain thermal modeling and experimental validation of 3D-ICs using integrated heaters and sensors [68].

For this review, the importance of test vehicles is methodological rather than merely historical. They make it possible to decouple specific questions that are otherwise confounded in full systems: the thermal effect of a single buried heater, the spreading penalty across a bonded layer, the temperature rise associated with a local hotspot geometry, or the degree to which a fine-grain BEOL-aware model reproduces the measured field. In other words, they bridge the gap between abstract material metrology and full-package validation. What they measure directly is sensor response under controlled heating; what they allow one to infer is whether a supposedly physics-grounded stack model reproduces the operative thermal pathways of a fabricated structure.

### E. WORKLOAD-BASED VALIDATION

Controlled heaters are essential, but they are not sufficient. Heterogeneous 3D ICs operate under time-varying, spatially heterogeneous workloads, and the resulting hotspot trajectories may differ substantially from those observed under synthetic or steady heating. That is why workload-based validation should be treated as a distinct stage rather than an optional add-on. Classic processor-thermal characterization work showed that different workloads produce markedly different thermal signatures and hotspot behavior, while more recent 3D-MPSoC evaluation frameworks use industry-standard benchmark suites such as SPEC CPU 2017 to drive thermal and dynamic thermal management studies under realistic application conditions [62], [68], [69].

In a heterogeneous 3D-IC setting, workload-based validation is the point at which the model must prove that it captures not only static thermal resistance, but also dynamic thermal causality: how hotspots appear, migrate, merge, and relax under realistic power traces. The key observables may come from on-die sensors, thermoreflectance snapshots, or package-level thermal data, but the inference target is broader: consistency between measured and predicted $T(x, y, t)$, peak temperature, hotspot location, and transient response under representative activity. This step is indispensable because a model that matches a thermal test vehicle under constant heating can still fail badly under realistic workload phasing, memory traffic, or heterogeneous compute behavior [62], [68], [69].

### F. INVERSE CALIBRATION AND PARAMETER IDENTIFICATION

The natural next step after measurement is inverse calibration. In heterogeneous 3D ICs, many of the parameters most needed by simulation: interfacial conductance, effective anisotropic conductivity, package-side boundary coefficients, compact-model resistances, and sometimes even power-map corrections are best obtained by solving an inverse problem rather than by direct lookup. Conceptually, the task is to determine a parameter vector $\theta$ such that the discrepancy between measured data and model prediction is minimized, often under regularization or Bayesian priors to control non-uniqueness. Recent FDTR inversion work and buried-interface thermoreflectance studies make this point explicit: the metrology signal becomes informative only after it is embedded in a model-based identification procedure [56], [66]

For review purposes, the main message is that calibration should be hierarchical. Interface properties should first be constrained by interface-sensitive metrology; layer or package parameters should then be calibrated against test structures or thermal fields; and only after that should compact or surrogate models be fitted to full-stack responses. If the order is reversed, high-level models can absorb lower-level errors and appear accurate for the wrong reasons. This is one reason why the validation workflow figure in your draft is so strong: it already treats calibration as an iterative loop between metrology, full-order simulation, reduced models, and updated parameter sets. That logic should be made explicit in this section. The principal risk, again, is identifiability. If too few observables are used, multiple parameter sets may explain the same data equally well, which leads to brittle extrapolation outside the calibration regime [60].

## VII. RELIABILITY, DESIGN CO-OPTIMIZATION, AND MITIGATION STRATEGIES

The practical value of multiscale thermal modeling is realized only when temperature prediction is translated into reliability assessment and actionable design decisions. In heterogeneous 3D ICs, this translation is especially important because the dominant thermal bottlenecks are closely coupled to aging, packaging stress, and runtime control. Elevated temperature not only reduces immediate performance margin but also accelerates long-term degradation mechanisms and narrows the feasible design space for floorplanning, power delivery, and cooling. Accordingly, thermal analysis in heterogeneous 3D stacks should be viewed not as a standalone signoff step, but as a co-optimization layer linking reliability, architecture, and mitigation strategy.

### A. RELIABILITY IMPLICATIONS OF ELEVATED TEMPERATURES

Elevated temperature accelerates several of the principal wear-out mechanisms in integrated circuits, including electromigration (EM), time-dependent dielectric breakdown (TDDB), bias-temperature-instability-related degradation, and thermal cycling fatigue [70]. In heterogeneous 3D ICs, these risks are amplified by the vertical concentration of power, the presence of repeated bonded interfaces, and the strong thermal coupling among tiers. As a result, local heating does not remain confined to one die or one interconnect level; instead, it can reshape current density, resistance, and stress fields across multiple layers, thereby increasing the effective aging rate of the stack [71].

Among these mechanisms, EM is particularly sensitive to the 3D integration context. The current path in advanced stacked systems traverses microbumps, redistribution structures, TSVs, and increasingly fine-pitch hybrid-bonded interconnects. This creates current crowding and temperature gradients that are more complex than in planar metallization, making EM assessment a genuinely three-dimensional problem rather than a simple extension of 2D wire reliability rules[71]. Likewise, thermo-mechanical reliability cannot be decoupled from thermal analysis. TSVs, bonded interfaces, Cu features, and adjacent dielectrics experience coefficient-of-thermal-expansion mismatch and nonuniform thermal loading, which can induce stress concentration, delamination, cracking, mobility variation, and timing degradation.

For heterogeneous stacks, the relevant temperature limit is also tier dependent. Logic, memory, analog, and power-oriented tiers do not necessarily share the same thermal envelope or the same dominant degradation mechanism. For example, memory layers may encounter temperature-dependent refresh and retention penalties before a neighboring logic layer reaches its own functional limit[72]. Therefore, a useful thermal model should not only estimate peak temperature, but also support safe-operating-area checks, lifetime-aware decreasing, and tier-specific thermal constraints tied to the governing reliability mode. In that sense, temperature prediction is most useful when it is turned into a reliability budget rather than treated as a scalar signoff number.

### B. THERMAL-AWARE FLOORPLANNING AND RUNTIME POWER MANAGEMENT

At design time, one of the most effective mitigation strategies remains thermal-aware physical planning. In stacked logic and chiplet systems, the placement of high-power blocks, the relative alignment of hotspots across tiers, and the available lateral spreading paths can materially alter peak temperature even when total power remains unchanged [17], [21], [73]. Thermal-aware floorplanning therefore attempts to distribute power more favorably, avoid vertically aligned hotspots, and exploit whitespace, tier assignment, or chiplet spacing to improve heat spreading before the package and cooling solution become fixed. In 2.5D systems, this logic extends naturally to chiplet placement; in true 3D stacks, it also includes vertical functional partitioning.

Thermal TSVs and other dedicated heat-conduction structures provide a second design lever. Properly placed thermal vias can lower hotspot temperature by creating additional vertical escape paths, but they also consume routing resources and may worsen stress or manufacturability if inserted without broader co-optimization [76], [74]. This means that thermal-via planning should not be treated as an isolated post-processing step. Instead, it should be co-optimized with signal/power TSV placement, floorplanning, and thermomechanical constraints so that a local temperature gain is not achieved at the expense of electrical or mechanical robustness.

At runtime, temperature management becomes a control problem. Dynamic voltage and frequency scaling (DVFS), clock and power gating, task migration, and temperature-aware memory-management policies are all more effective when driven by compact thermal models rather than by threshold-only sensing [72], [75]. In stacked systems, this is particularly important because thermal coupling across tiers means that throttling one layer can indirectly cool, or sometimes thermally redistribute load onto, another. Thus, runtime controllers for heterogeneous 3D ICs should be tier aware and coupling aware, not purely local. This is one of the clearest examples of why compact thermal and electrothermal models remain essential even when higher-fidelity field solvers are available offline.

### C. MATERIALS AND PACKAGING INNOVATIONS
Thermal mitigation in heterogeneous 3D ICs increasingly depends on materials and packaging co-design. At the material level, electrically insulating layers with improved thermal conductivity can reduce BEOL and interposer spreading resistance, thereby easing one of the most persistent bottlenecks in vertical integration. At the bonding level, fine-pitch hybrid-bonding schemes can reduce some of the geometric and parasitic penalties associated with conventional microbump-based interconnects, but they also make local interface quality, Cu/dielectric patterning, and buried-contact integrity more central to thermal performance [76]. In other words, newer packaging technologies may reduce one class of thermal bottleneck while making another more model sensitive.

At the cooling level, embedded liquid and microfluidic approaches have attracted sustained attention because they shorten the distance between the heat source and the coolant, which is especially attractive in stacks where heat generated in interior tiers is difficult to remove through a conventional top-side heat sink alone [77]. The key point for this review is not to rank these strategies universally, but to emphasize that each one changes the modeling problem. High-thermal-conductivity insulators change the effective anisotropic tensor used in continuum solvers. Hybrid bonding changes the relevant interface model and the importance of buried metrology. Liquid cooling changes the boundary-condition structure and often the dominant timescales. Thus, mitigation strategies and thermal models must be co-developed rather than evaluated independently.

### VIII. OPEN CHALLENGES AND FUTURE DIRECTIONS
Despite substantial progress, several challenges continue to limit predictive thermal analysis in heterogeneous 3D ICs. Some of these challenges arise from scaling itself, especially the move toward finer-pitch bonding and higher local power density. Others arise from methodology: incomplete multiphysics coupling, lack of standardized interface data, and uncertainty that is often ignored once a model is reduced or calibrated. A final group arises from new application domains, including cryogenic quantum-classical systems and runtime digital twins. The following issues appear especially consequential.

### A. MODELING AT SUB-10 μM HYBRID-BOND PITCH
As die-to-die pitch approaches the single-digit-micrometer regime, simple homogeneous-interface abstractions become increasingly fragile. In such stacks, the bonded region is no longer well described by a single averaged layer unless the averaging procedure itself is layout aware. Instead, local Cu/dielectric pattern density, alignment, voiding, and buried-contact quality can materially affect both electrical and thermal behavior [76]. The resulting challenge is not only higher fidelity, but reducibility: detailed layout-sensitive transport must still be translated into compact quantities usable by stack-level models.

For thermal modeling, this means that future workflows will likely require tighter coupling between detailed interconnect-scale analysis and higher-level solvers. Buried-interface metrology such as FDTR/TDTR can help constrain those models, but it does not eliminate the need for careful reduction and uncertainty assessment [68], [69]. The open problem is therefore to build models that remain physically faithful to sub-10-μm bonded layouts while still being fast enough for architecture and package co-design.

### B. CO-SIMULATION OF POWER DELIVERY, SIGNAL INTEGRITY, AND THERMAL BEHAVIOR
In advanced heterogeneous stacks, electrical and thermal behavior increasingly shape one another. Dense power-delivery and signal-distribution networks change current crowding, Joule heating, and local spreading paths; conversely, temperature and stress perturb resistance, timing, and in some cases interconnect reliability. Existing work on signal/power/thermal co-design, full-chip electrothermal extraction for face-to-face bonded 3D ICs, and electrical–thermal–mechanical modeling of TSV-aware clock networks demonstrates the importance of this integration, but robust full-stack flows for heterogeneous chiplet packages and fine-pitch bonded assemblies remain limited [36], [83], [84].

The challenge is partly algorithmic and partly organizational. Different tools often operate on different abstractions, meshes, and design hierarchies. As a result, it remains difficult to propagate electrothermal or thermo-mechanical effects consistently from detailed interconnect structures up to chiplet- or package-level decisions. This is likely to become even more important as advanced stacks place tighter constraints on power delivery, current density, and interface-aware cooling.

### C. STANDARDIZED TBR LIBRARIES AND UNCERTAINTY QUANTIFICATION
Interfacial thermal resistance remains one of the most consequential and least standardized inputs to heterogeneous 3D IC thermal analysis. Reported values can differ substantially because they depend on interface chemistry, roughness, pressure, defectivity, film thickness, transducer assumptions, and inversion methodology [8], [68], [69]. In other words, the field does not merely lack more measurements; it lacks sufficiently standardized metadata-rich interface libraries that define what was measured, how it was inferred, and what uncertainty should accompany the result.

The related challenge is uncertainty quantification. Even when a nominal $G_{int}$, $k$, or $h$ value is identified, the uncertainty attached to that value is rarely propagated into the final chip/package thermal prediction. Yet lower-scale uncertainty does not disappear when converted into an effective parameter; it is simply carried upward in transformed form [74], [75]. Future heterogeneous 3D IC workflows should therefore report not only nominal temperature fields, but also confidence bounds on peak temperature, inter-tier gradients, and reliability-relevant exposure metrics.

### D. CRYOGENIC AND QUANTUM-CLASSICAL HETEROGENEOUS SYSTEMS

A further frontier lies in cryogenic heterogeneous systems, particularly those combining cryo-CMOS control/readout electronics with quantum or superconducting devices. At cryogenic temperatures, thermal modeling must account for strongly temperature-dependent material properties, limited refrigeration power, self-heating at cold stages, and interfacial thermal boundary resistance of the Kapitza type [78], [79], [80]. These effects differ substantially from the room-temperature assumptions built into most existing compact and package-scale thermal tools.

The motivation is clear: bringing control electronics closer to qubits can improve scalability and wiring complexity, but it also introduces tight thermal budgets that may be incompatible with room-temperature modeling assumptions. This creates an opportunity for new multiphysics workflows that couple cryogenic transport, interface resistance, and circuit operation across large temperature gradients. For your paper, this topic works best as a forward-looking extension case rather than a central pillar, but it is still worth retaining because it highlights where the current modeling ecosystem is least prepared.

### E. MACHINE-LEARNING ENABLED RUNTINE THERMAL PREDICTION

Finally, future heterogeneous 3D systems are likely to rely more heavily on runtime thermal prediction using ML-accelerated surrogates, operator-learning models, or hybrid physics-plus-data digital twins [49]–[55]. The attraction is obvious: once trained, these models can deliver temperature fields or hotspot metrics far faster than repeated numerical PDE solves, making them suitable for inner-loop optimization, adaptive control, and online monitoring.

The main challenge is no longer raw prediction speed, but robustness under shift. Workload patterns, package boundary conditions, silicon aging, and manufacturing variation can all move the operating point outside the training manifold. As a result, future runtime thermal models must become more uncertainty aware, more explainable, and more tightly tied to physics-based reference models and sensor-based recalibration. The long-term opportunity is a hierarchical digital-twin workflow in which validated full-order models, compact models, and learned surrogates are updated continuously rather than used as fixed, disconnected artifacts.

## IX. CONCLUSION

Heterogeneous 3D integration offers a powerful path beyond conventional monolithic scaling by enabling dense co-integration of logic, memory, and other specialized functions within a single package. However, the same architectural features that make these systems attractive vertical stacking, fine-pitch die-to-die interconnects, heterogeneous materials, and compact package integration also make thermal behavior a first-order design constraint. Heat generation is highly nonuniform, heat-removal paths are restricted, and repeated interfaces introduce additional resistance and uncertainty. As a result, thermal analysis in heterogeneous 3D ICs cannot be treated as a secondary packaging concern; it must be incorporated as a central part of architecture, physical design, and reliability evaluation.

This review has examined the governing thermal physics of heterogeneous 3D ICs, including thin-film and size effects, thermal boundary resistance, anisotropic heat transport in BEOL and interconnect structures, and electrothermal and thermomechanical coupling. It has also organized the modeling landscape into a multiscale hierarchy spanning first-principles and molecular-level approaches, mesoscopic phonon transport models, continuum FEM/FVM and Green's-function solvers, compact electrothermal networks, reduced-order models, and emerging ML/PINN-based surrogates. A central conclusion is that no single modeling framework is sufficient across all scales or all design tasks. Instead, predictive analysis depends on physically consistent cross-scale workflows in which lower-scale physics is translated into effective parameters, higher-level models are calibrated against measurements and reduced or learned models are deployed only within validated domains of use.

Across representative stack classes such as logic–HBM systems, stacked-logic assemblies, and chiplet/interposer-based packages, the most reliable modeling strategy today is a calibrated multilevel workflow: interface- and material-sensitive parameters are obtained from fine-scale simulation or metrology, full-stack temperature fields are resolved using continuum solvers, and compact or reduced-order models are then derived for design exploration and runtime management. Experimental validation is therefore not an afterthought but a structural pillar of the modeling chain. On-die sensors, thermoreflectance, Raman thermography, TDTR/FDTR, dedicated test vehicles, and workload-driven measurements each constrain different parts of the model, while inverse calibration and uncertainty propagation determine how confidently thermal predictions can be used in design decisions.

Several open challenges remain. Future heterogeneous 3D systems will require better thermal abstractions for sub-10-μm hybrid-bonded interfaces, tighter co-simulation of power delivery, signal integrity, thermal transport, and mechanical stress, more standardized and uncertainty-aware TBR libraries, and more trustworthy runtime surrogates for adaptive thermal management. Addressing these challenges will be essential for turning multiscale thermal modeling from a powerful analysis tool into a robust design methodology. Ultimately, the success of heterogeneous 3D integration will depend not only on advances in packaging and bonding technology, but also on the ability to predict, validate, and co-optimize thermal behavior across scales with sufficient fidelity, speed, and confidence.